\documentclass[a4paper]{article}

\usepackage{INTERSPEECH2022}

\usepackage[dvipsnames]{xcolor}
\usepackage{bm}

\usepackage[colorlinks,citecolor=red,urlcolor=blue,bookmarks=false,hypertexnames=true]{hyperref}

\title{End-to-end speech recognition modeling from de-identified data}
\name{Martin Flechl, Shou-Chun Yin, Junho Park, Peter Skala}

\address{
  Nuance Communications Inc., Burlington, USA
}
\email{martin.flechl@nuance.com}

\begin{document}

\maketitle
\begin{abstract}
De-identification of data used for automatic speech recognition modeling is a critical component in protecting privacy, especially in the medical domain.
%De-identification of training data for automated speech recognition has become a necessity for privacy reasons, especially in the medical domain. 
However, simply removing all personally identifiable information (PII) from end-to-end model training data leads to a significant performance degradation in particular for the 
recognition of names, dates, locations, and words from similar categories. 

We propose and evaluate a two-step method for partially recovering this loss. First, PII is identified, and each occurrence is replaced with a 
random word sequence of the same category. Then, corresponding audio is produced via text-to-speech or by splicing together matching audio fragments extracted from the corpus.
%We propose and evaluate the following 
%method for partially recovering this loss: First, PII is identified, and each occurrence is replaced with a random 
%word sequence of the same category. Then, corresponding audio is produced in different ways: By using text-to-speech, or by splicing 
%together matching fragments of audio extracted from the corpus.
%%, or a combination of both. 
These artificial audio/label pairs, together with 
speaker turns from the original data without PII, are used to train models. We evaluate the performance of this method on in-house data of medical conversations and observe a recovery 
of almost the entire performance degradation in the general word error rate while still maintaining a strong diarization performance. 
Our main focus is the improvement of recall and precision in the recognition 
of PII-related words. Depending on the PII category, between $50\% - 90\%$ of the performance degradation can be recovered using our proposed method.
\end{abstract}

\section{Introduction}
Two of the important ingredients to a performant automatic speech recognition (ASR) system are the quality and quantity of the training data. Special care is needed to avoid any 
mismatch between the data used for model training and for model application in the field, for example in terms of acoustic conditions, speaker 
variability and vocabulary usage. This is challenged by requirements to de-identify training data~\cite{hipaa}: In its simplest form, this would imply to 
simply remove all PII from both the audio and label part of the training corpus. However, a model which does not see any person names, dates, 
or locations during training is likely not able to reliably recognize such entities in the field. 
While the impact on overall metrics such as word error rate (WER) might still be small such a system would nonetheless not be suited for most practical applications.
For hybrid ASR models, the acoustic part of PII data 
(phonemes, with or without context) is usually covered by audio from a non-PII context and the problem can be efficiently dealt with by only manipulating text 
in order to make sure the PII tokens are adequately represented in the language model (LM). For end-to-end ASR modeling without LM fusion~\cite{lmfusion}, however, 
in addition to text manipulation the arguably more demanding generation of audio is required in some form.

We address this problem by enriching the training data with words from PII categories (for example, names or dates) while minimizing the impact on the data 
in any other respect. Specifically, we first identify PII in the audio and text labels of the training data. Then, instead of just removing it, 
we replace each occurrence with a word sequence of the same category. For the text labels, this is straight-forward. The challenge lies in providing 
audio corresponding to these inserted text labels. For this task, we compare different configurations employing text-to-speech (TTS) or matching fragments 
extracted from the corpus.

Utterances which do not contain any PII to start with are combined with utterances de-identified and enriched in the described 
way to train ASR models. We use state-of-the-art end-to-end models based on the Conformer-Transducer architecture~\cite{conformer} 
to evaluate the performance degradation caused by removing all PII and how much of this loss can be recovered.

In the past few years, neural TTS models have managed to produce results comparable to human speech~\cite{tts0,tts1,tts2,tts3,tts4}. 
TTS has many applications related to ASR and is an obvious choice to generate audio counterparts for unpaired text for end-to-end ASR training~\cite{deliberation_jatd}. 
Reusing audio snippets from a corpus (``splicing'') has previously been used in the context of domain adaptation as a cheap and scalable way to produce specific training 
data without additional data collection~\cite{splicing}. End-to-end ASR models have become the state-of-the-art in recent years in terms of reducing 
recognition error rates~\cite{las,transformer,transformertransducer}. However, they also introduce a new set of challenges compared to hybrid ASR models. 
In this paper, we apply the established methods of TTS and splicing to reduce the negative impact of training data de-identification on end-to-end ASR models 
and evaluate the results of a few distinct strategies.

\section{De-identification}
To protect privacy and in some cases meet contractual or regulatory obligations, eliminating PII from collected data has become an important step across many fields. 
Methods for masking sensitive text elements and their impact on machine learning tasks have been now studied for almost twenty years for the general case~\cite{deid1} 
but also specifically in the healthcare domain~\cite{deid3}. Several methods have been suggested to de-identify health records~\cite{deid4} and patient notes~\cite{deid5, deid6}.
An alternative way of preserving privacy is via federated learning where sensitive data is exclusively processed on the client side~\cite{zhu22_interspeech}.

%Due to regulatory measures, contractual obligations and in general a heightened sensibility for privacy concerns, eliminating PII from collected data has 
%become a necessary or at least desirable strategy for many fields. 
In this paper, we deal with ASR for medical conversations which is arguably one of the 
most sensitive applications in this respect. In the following, de-identification is understood as the process of removing PII from speech, i.e., from 
spoken text (audio) or transcribed text. Examples are patient names or birth dates. Speaker anonymization, i.e., eliminating biometric attributes, 
is not subject of this paper.

The de-identification process starts with manual annotation of the transcribed text. PII is tagged and given one of roughly 30 labels. 
The audio interval containing the PII is found via forced alignment of the transcription to the audio. 
Then the PII is removed from the transcription and the waveform of the corresponding audio interval is replaced by silence. The result 
of this process is new audio and new text in which blanks replace the original PII.

\section{Strategies for performance recovery}
\begin{figure}[ht]
  \centering
  \includegraphics[width=\linewidth]{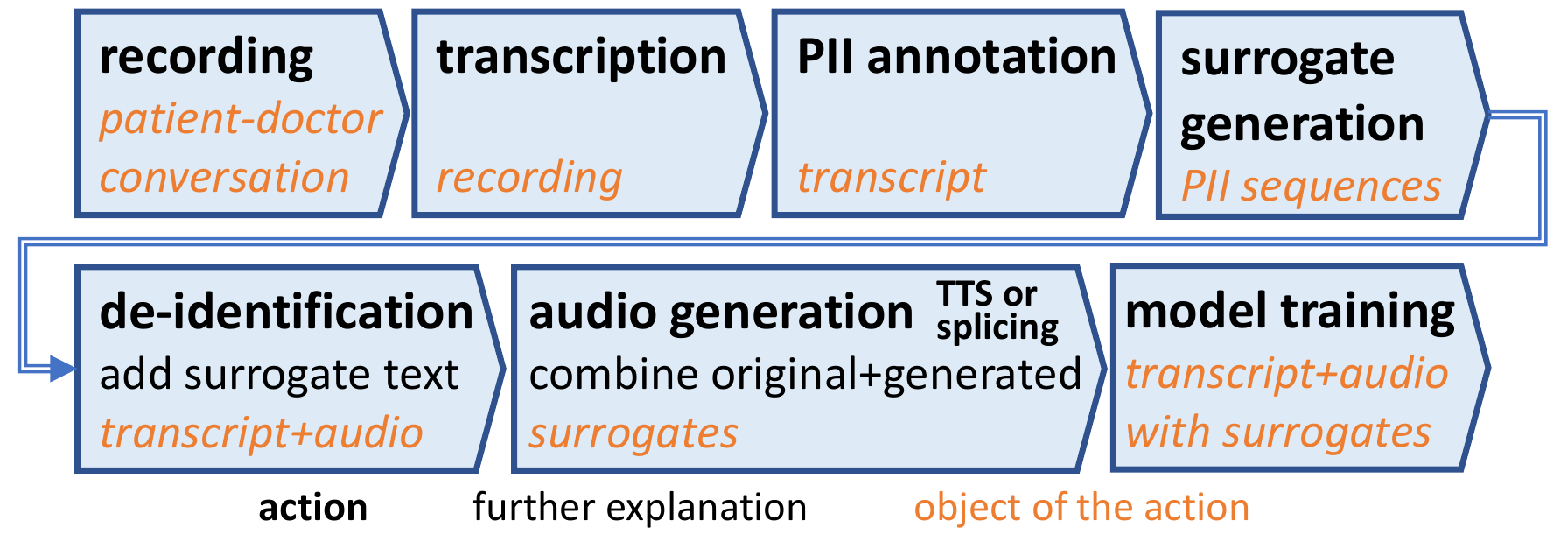}
  \caption{Method workflow. From the recording of the conversation to the model training with data enhanced by de-identification mitigation strategies.}
  \label{fig:method}
\end{figure}
% 1. recording
% 2. transcription (auto/manual corr.)
% 3. PII annotation (auto/manual corr.)
% 4. surrogate generation
% 5. de-identification (text, audio)
% 6. audio generation for surrogates (TTS / splicing) + stitching
% 7. model training on new training data
%
The performance of models trained with identified data is the best-case scenario and defines the upper baseline. Using training data 
after simply removing all PII is the corresponding worst-case scenario and lower baseline. Past investigations have shown that 
such a removal leads to a significant drop in performance for natural language processing tasks while suitably replacing sensitive words 
is more robust~\cite{adelani20_interspeech} in that respect. The goal of the methods described in the following 
is to recover as much as possible from the performance gap between lower and upper baseline without using PII. In all cases, the original 
PII text is replaced by a so-called surrogate. 
Different methods and strategies are then employed to also populate the corresponding audio. 
The workflow for the whole process is illustrated in Figure~\ref{fig:method} and explained in the following.

\subsection{Surrogates}
Surrogates are generated in a pseudorandom and consistent manner: each PII phrase from a given category is replaced by 
a surrogate from the same category.
This means that e.g., ``Michael'' may be replaced by ``John'', ``New York'' by ``Los Angeles'',
``6/3 2021'' by ``7/4 2019'' and ``April 7, 2020'' by ``first of May, 2021''. 
This completes the treatment of the text labels of the corpus; two different methods to deal 
with the audio part are described in the following.

\subsection{Text-to-speech}
\begin{figure}[ht]
  \centering
  \includegraphics[width=\linewidth]{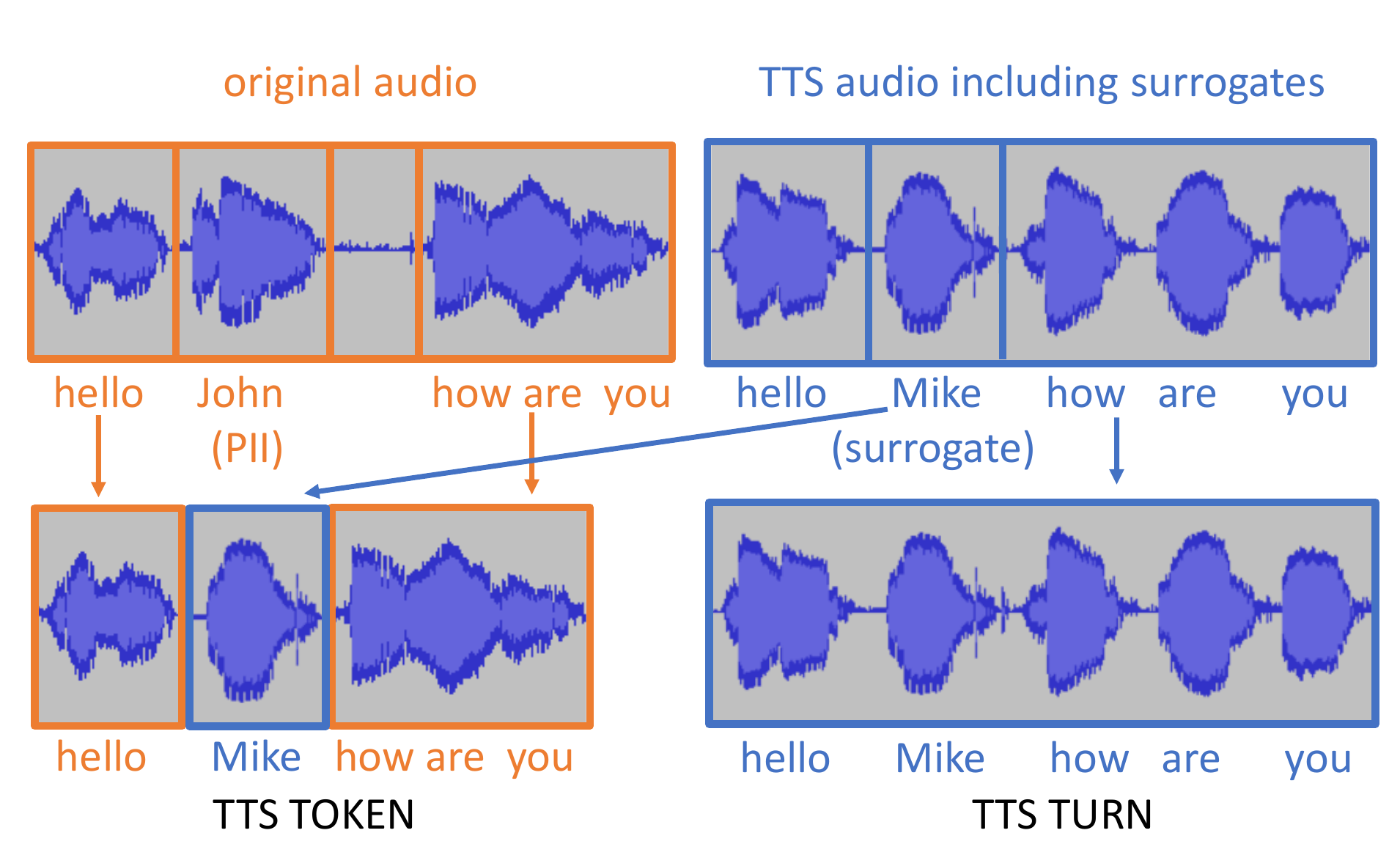}
  \caption{Audio generation by combining original speech and TTS, or replacing original speech with TTS.}
  \label{fig:tts}
\end{figure}
The first method uses TTS to generate audio corresponding to the surrogate. Two different strategies are compared, as illustrated in Figure~\ref{fig:tts}.
\begin{itemize}
\item
\textbf{\textit{TTS TOKEN}}: Only the tokens of the surrogate are replaced by TTS. The complete speaker turn is then stitched together from parts of the original audio and TTS parts.
\item
\textbf{\textit{TTS TURN}}: Any turn containing surrogates is entirely replaced by TTS. No stitching takes place; each turn is either entirely from the original audio or generated via TTS.
\end{itemize}
For the experimental studies presented in this paper, Cerence TTS~\cite{cerence} is used to synthesize clean speech audio with one of eleven English speaker voices 
(four male, seven female) where we choose one randomly for each turn. 
Before the TTS audio snippet is concatenated with the original audio, it is processed in the following ways to smoothen the transition: the volume is adjusted 
to the same level as the original conversation, and silence is trimmed at the beginning and at the end of the audio snippet.

\subsection{Splicing using audio from the corpus}
\begin{figure}[ht]
  \centering
  \includegraphics[width=0.96\linewidth]{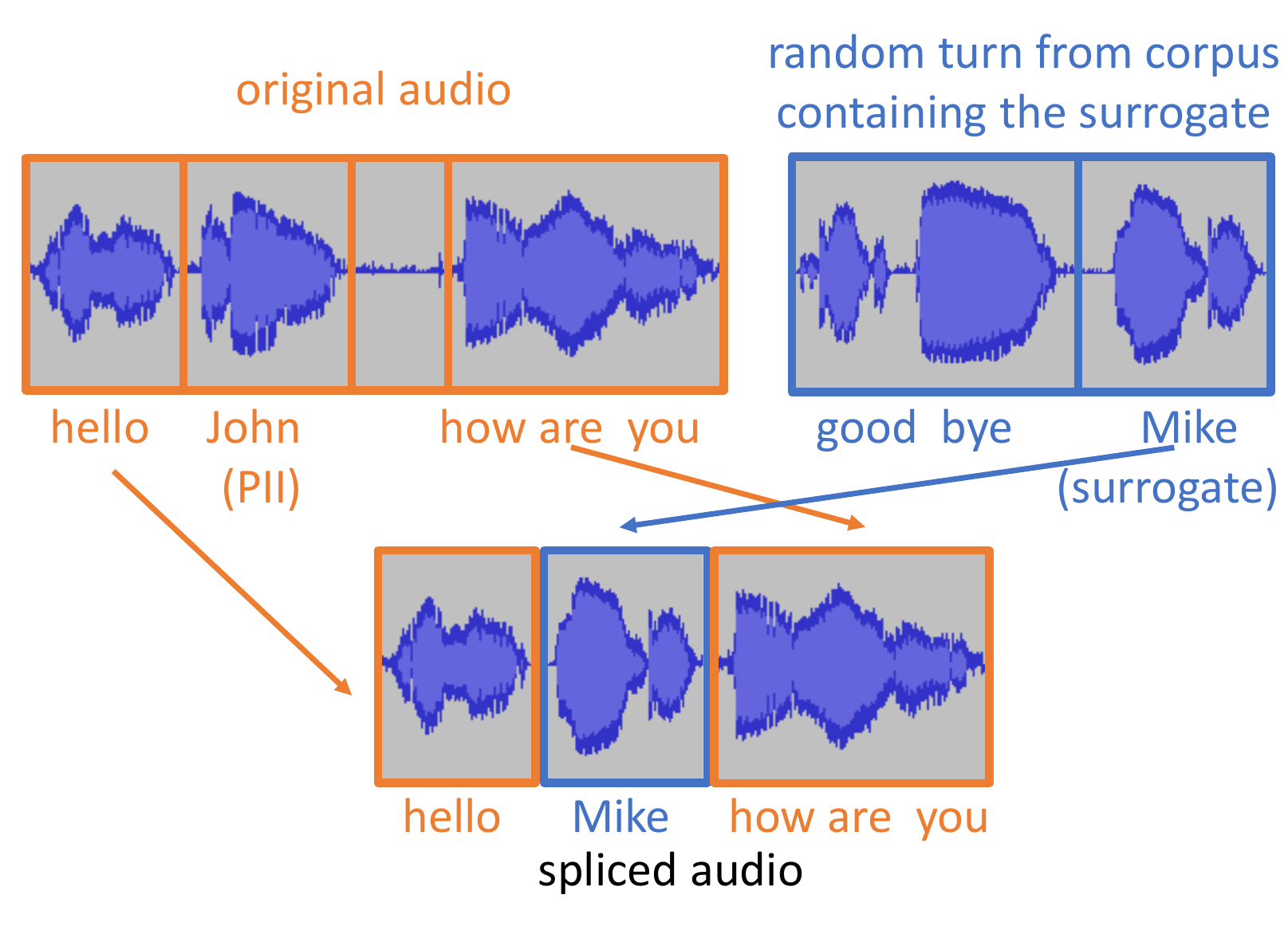}
  \caption{Audio generation by splicing original speech and snippets from the corpus.}
  \label{fig:splicing}
\end{figure}
The second method is based on a strategy developed originally to adapt an existing model to a new domain without collecting additional data~\cite{splicing}. 
The surrogate tokens are searched for in the corpus. They are then extracted using the token-level timing information obtained from a forced alignment of transcription and audio. 
Two different strategies are evaluated:
\begin{itemize}
\item
\textbf{speaker-dependent \textit{SD}}: We only use corpus audio from the speaker of that turn; if such a spoken word is unavailable then the parts of the turn containing PII tokens 
are dropped.
%We only use audio from our corpus which are from the speaker of the original turn.
\item
\textbf{speaker-preferred \textit{SP}}: Same as \textit{SD}, but with fallback to using audio from any other speaker if the required word cannot be found in the corpus for the speaker of that turn.
\end{itemize}
The complete turn is then spliced together from parts of the original audio and the extracted audio snippet, see Figure~\ref{fig:splicing}. 
In the speaker-preferred case, the final turn potentially 
consists of audio snippets from different speakers. In both cases acoustic conditions may differ (e.g., different level of background noise or distance to the microphone). 
The speaker-dependent strategy is more consistent and a priori more compatible with the desired modeling of the diarization. 
However, it has been argued~\cite{splicing} 
that end-to-end models are particularly robust with respect to discontinuities at the transition between words. We address this question empirically by comparing to the 
speaker-preferred strategy. The latter has the advantage of a higher efficiency: if the search for a specific word is restricted to one speaker it is more likely it cannot 
be found in the corpus and hence the turn (or at least part of the turn) cannot be used for training. 
If more than one matching token is found then a random choice is made.

\section{Experiments}
The training and test data, model architecture and other aspects of the training setup used to evaluate the performance of the different strategies are described in the following.

\subsection{Data}
The training data consist of a subset of about $1000$ hours of an in-house dataset containing medical conversations conducted in English covering about twenty different medical specialties. 
The recording device is carried by the doctor and hence the conversation typically consists of a mixture of near-field speech from a doctor and far-field speech from patient, 
caretakers, other doctors, and nurses. The audio is characterized by varying background noise and reverberation conditions. The test data are not de-identified but otherwise 
have the same characteristics and cover about $73$ hours. The recordings are manually transcribed and PII sequences are annotated afterwards.

In total, the annotated PII makes up $1.4\%$ of the tokens in our training data and $2.1\%$ of the total audio time. About $81\%$ of all PII tokens are 
uttered by the doctor. 
% , mostly patient names, birth dates and medical record numbers. 
For this reason the replaced audio mainly consists of near-field speech which is 
easier to generate artificially compared to far-field speech. Of all PII tokens, $45 \%$ are 
labeled as dates and $30 \%$ as patient or doctor names. The rest are numbers (medical records, phone numbers, age, etc) and names of places, hospitals or other organizations. 
Person names and dates are essentially entirely removed while numbers are only suppressed to a variable degree as they also occur in non-PII contexts.

% ENCODER
% https://git.labs.nuance.com/research_tools/dante/-/blob/872f7d98929f517f2920b0d12cd92b9e09625da7/dante/layers/conformer.py#L263
% input: wave file
% 1. FT, 10ms steps, 32ms windows
% 2. MFCC features, 64 dim
% 3. CMVN (Cepstral mean and variance normalization)
% 4. BatchNormalization
% 5. SpecAugment, 10 time masks (p=0.05, max time/mask ratio) and 2 frequency masks
% 6. two layers of 2D convolutions, 64 filters, kernel 3x3, stride 2x2 => convolution subsampling to 40 ms windows; w/ batch norm
% 7. dense layer, 512 units, dropout 0.1
% 8. 18 conformer blocks: FF1(2048, dropout 0.1) + MHSA(8, dropout 0.1) + Conv(kernel 7, dropout 0.1) + FF2(2048, dropout 0.1) + LayerNorm
% 9. FF(512)
%
% PREDICTION NETWORK
% https://git.labs.nuance.com/research_tools/dante/-/blob/872f7d98929f517f2920b0d12cd92b9e09625da7/dante/layers/transformer.py#L200
% input: previous outputs, lookback=50
% 1. 1 transformer decoder block: MHSA (8, dropout 0.2, res. conn.) + PWFF(layer norm + FF1(512, dropout 0.1) + FF2(1024, dropout 0.1) + layer norm)        [mhxa skip!]
% 2. FF(512)
%
% JOINT NETWORK
% https://git.labs.nuance.com/research_tools/dante/-/blob/872f7d98929f517f2920b0d12cd92b9e09625da7/dante/layers/rnn_transducer.py#L166
% input: concat of encoder and prediction network
% 1. FF(256)
% 2. FF(voc size=1024, dropout 0.1)
% 3. softmax
\subsection{Experimental setup}
We train end-to-end ASR models which in addition jointly predict diarization~\cite{2joint} and auto-punctuation and are based on the Conformer-Transducer architecture~\cite{conformer}. 
The inputs are 
pairs of wave files and sequences of word-pieces, created using a sentence piece model~\cite{spm} 
with a vocabulary size of $1024$. The encoder extracts $64$-dimensional Mel Frequency Cepstral Coefficents features from $32$\,ms windows in $10$\,ms steps. 
During training SpecAugment~\cite{specaugment} 
%with ten time masks and two frequency masks 
is applied. Two convolutional layers are used to subsample the features by a total factor of four, resulting in $40$\,ms frames. 
After a dense layer and a 
dropout layer, eighteen conformer blocks constitute the workhorse of the encoder. The macaron-style conformer blocks~\cite{conformer} combine a feed-forward network, multi-head attention, 
a convolutional layer, another feed-forward network and layer normalization. After the conformer blocks, a feed-forward network with $512$ units produces the encoder output.

The prediction network is fed by previous model outputs and consists of a multi-head self-attention layer and three feed-forward network layers with 512 output units. The joint network 
concatenates the encoder and prediction network outputs and consists of two feed-forward network layers and a dropout layer. A softmax layer produces the final model output. In total, 
the model features 113 million trainable parameters.

All models are trained on six Tesla V100 GPUs with TensorFlow 2~\cite{tf} for 80~epochs which is sufficient for the training to converge and 
takes about four days for the $1000$h training data subset used. 
Recognition performance is evaluated using an 
average of the last fifteen model 
checkpoints. Where we report results on continuing model training based on initializing with the final weights of another model training, 
this implies an additional training for fifteen epochs. We report the usual word error rate (WER) but also the word 
diarization error rate (WDER)~\cite{2joint}
which measures the fraction of words not assigned to the correct speaker (``doctor'' or ``other'') to assess if any of the methods confuses the diarization 
performance of the model. However, the main focus is on tokens which have a significantly reduced frequency in the training corpus after de-identification if no recovery methods 
are applied. These tokens are either chosen by category (\textit{numbers}, \textit{dates}, and \textit{names}) or agnostically directly based on their frequency ratio in the training 
corpus before and after de-identification, looking at ratio bands below $10\%$ and between $10\%-20\%$.

The following models are trained:
\begin{itemize}
\item
\textit{baseline ID} (B1): identified training data, i.e., no de-identification has taken place.
\item
\textit{baseline TURN} (B2): all speaker turns with PII have been removed.
\item
\textit{baseline TOKEN} (B3): all PII tokens have been removed (and the rest of each turn is kept).
\item
\textit{TTS TOKEN} (T1): PII tokens have been replaced by surrogates, and for the audio part, PII tokens are generated using TTS.
\item
\textit{TTS TURN} (T2): PII tokens have been replaced by surrogates, and for the audio part the whole turn is generated using TTS.
\item
\textit{spliced SD} (S1): PII tokens have been replaced by surrogates, and for the audio part matching snippets from the same speaker are used if available; otherwise the turn is skipped.
\item
\textit{spliced SP} (S2): PII tokens have been replaced by surrogates, and for the audio part matching snippets from preferably the same speaker (with fallback to any speaker) are used 
if available; otherwise the turn is skipped.
\item
\textit{spliced SDc} (S3): same as \textit{spliced SD}, but initializing the model with \textit{baseline TOKEN} weights instead of starting from scratch.
\item
\textit{spliced SPc} (S4): same as \textit{spliced SP}, but initializing the model with \textit{baseline TOKEN} weights instead of starting from scratch.
\end{itemize}

\subsection{Results}
The resulting word error rates and word diarization error rates are shown in Table~\ref{tab:results_wer}. B1 serves as a best-case baseline, with all information present; B2 and B3 
constitute the worst-case baselines. 
The first thing to note is that the WER gap between upper and lower baselines is relatively small. This is not surprising considering that only about $1.4 \%$ of our tokens are 
PII-tagged and hence the impact on general performance numbers is expected to be of that order of magnitude. Nonetheless it is desirable to close this gap. 
As can be seen, the best models working with de-identified data and our mitigation methods indeed do not show a significant difference in WER with respect to the best-case baseline. 

The speaker-preferred splicing method gives significantly better results 
than the speaker-dependent one: the reason is the considerably larger percentage of successfully spliced turns of $66\%$ for \textit{spliced SP} compared to $17\%$ 
for \textit{spliced SD}. 
It is also worth noting that if the weights of the splicing models are initialized from a de-identified baseline model, results significantly improve. 
For TTS, the \textit{TOKEN} method which maximizes the amount of original audio performs comparably to the \textit{TURN} method which minimizes discontinuities in the audio. 
The WER and WDER are similar to the de-identification baseline.

Concerning diarization, as expected de-identification does not directly impact the performance and WDER numbers do not significantly differ between the baselines. 
More importantly, the diarization performance does not suffer from either TTS or splicing methods. This provides evidence that our models are robust with respect 
to the artificial transitions which these methods add to the audio part of our training data. 
Note in particular that the diarization performance does not degrade for the speaker-preferred with respect to the speaker-dependent splicing method: even though 
most SP-spliced speaker turns contain parts uttered by two or more speakers, the model does not get confused. 
In our data, about two thirds of all uttered words are from the doctor. 
More precisely, the naive baseline of attributing all turns to the 
dominant speaker has a WDER of $32.3\%$.

%TODO
%
% >>> tot=35811+18366
% >>> sdpref=35811/tot
% >>> sd=9227/tot
% >>> sd2=9227/(9227+26588)
% >>> sdpref, sd, sd2
% (0.6610000553740517, 0.17031212507152482, 0.25762948485271536)
\begin{table}[th]
  \caption{Performance in terms of WER and WDER of the models described in the text body with respect to baseline models.
}
  \label{tab:results_wer}
  \centering
  \begin{tabular}{ l l | r r r }
%    \toprule
    \multicolumn{1}{c}{\textbf{id}} & \multicolumn{1}{c|}{\textbf{model}} & \multicolumn{1}{c}{\textbf{WER}} & \multicolumn{1}{c}{\textbf{WDER}} \\
\hline
%    \midrule
    B1                              & \textit{baseline ID}                &  \color{blue}         \bf{11.8}  & \color{blue}           \bf{3.7}  \\
    B2                              & \textit{baseline TURN}              &                          $12.3$  &                           $3.8$  \\
    B3                              & \textit{baseline TOKEN}             &  \color{Red}          \bf{12.1}  & \color{Red}            \bf{3.7}  \\
\hline
    T1                              & \textit{TTS TOKEN}                  &                          $12.2$  & \color{ForestGreen}    \bf{3.6}  \\
    T2                              & \textit{TTS TURN}                   &                          $12.2$  &                           $3.8$  \\
\hline
    S1                              & \textit{spliced SD}                 &                          $12.4$  &                           $3.9$  \\
    S2                              & \textit{spliced SP}                 &                          $12.0$  &                           $3.8$  \\
    S3                              & \textit{spliced SDc}                &                          $11.9$  &                           $4.4$  \\
    S4                              & \textit{spliced SPc}                &  \color{ForestGreen}  \bf{11.8}  &                           $4.0$  \\
%    \bottomrule
  \end{tabular}
\end{table}
The main goal of this paper is to improve the recognition of PII-like words. 
Table~\ref{tab:results_f1} shows corresponding results for the F1 scores of any \textit{numbers} (e.g., two, ten, hundred), \textit{date}-related words (months or days of the week) 
and common English first \textit{names} where in all cases we exclude words that frequently occur with a different 
meaning (e.g., ``one'' from numbers, ``May'' from months and names). We look at first names because they have a significant overlap between training and test data. 
Surnames and uncommon first names of the test set mostly do not occur in the training data even before de-identification: while this is a general challenge for ASR 
it is not a problem related to de-identification and hence not addressed here. In addition to these categories, we also aggregate words based on how strongly 
they are suppressed by de-identification.

Both de-identified baseline models (B2, B3) show a strong degradation, 
most noticeably for \textit{names} where F1 scores drop from $77\%$ to below $23\%$. 
The TTS and splicing models recover a significant portion of the F1 score drop. Typically, they have similar precision to the \textit{baseline ID} model but slightly worse recall. 
The splicing models perform poorly with respect to \textit{names} which is explained by the fact that the splicing efficiency is much lower for names compared to numbers, days of the 
week or months. This is not the case for TTS which performs best for \textit{names} and in general tokens which are strongly suppressed in the training corpus due do de-identification. 
The recognition of \textit{numbers} is decent even when training only with de-identified data since numbers occur also outside of PII contexts and are thus not entirely suppressed 
in the training data. However, note that even here the ($1-$F1) error rate rises from $3 \%$ (B1) to $5 \%$ (B2, B3) after removing PII from the training data and our methods 
recover about half of that loss.
\begin{table}[th]
  \caption{
Performance in terms of F1 scores for PII-related words of the models described in the text body with respect to baseline models.
``Dates'' are days of the week or months; ``names'' cover the most common English first names. 
``r10'' are tokens with a frequency ratio in the training corpus after versus before de-identifaction of less than 10$\%$, ``r20'' of between 10$\%$ and 20$\%$. 
}
  \label{tab:results_f1}
  \centering
  \begin{tabular}{ l | r r r r r r }
%    \toprule
                                     &  \multicolumn{5}{|c}{\textbf{F1 score}} \\
    \multicolumn{1}{c|}{\textbf{id}} & \multicolumn{1}{c}{\textbf{numbers}} & \multicolumn{1}{c}{\textbf{dates}} & \multicolumn{1}{c}{\textbf{names}} 
                                     & \multicolumn{1}{c}{\textbf{r10}}     & \multicolumn{1}{c}{\textbf{r20}} \\
\hline
    B1                               & \color{blue} \bf{96.9} & \color{blue} \bf{96.4} & \color{blue} \bf{77.0} & \color{blue} \bf{85.5} & \color{blue} \bf{86.9}  \\
    B2                               & $94.7$                 & $51.7$                 & $22.7$                 & $43.6$                 & $74.6$                  \\
    B3                               & \color{Red}  \bf{95.1} & \color{Red}  \bf{59.9} & \color{Red}  \bf{21.4} & \color{Red}  \bf{48.2} & \color{Red}  \bf{75.5}  \\
\hline
    T1                               & $95.0$                 & $83.0$                 & $40.8$                 & $64.7$                 & $79.6$                  \\
    T2                               & $95.2$                 & $91.3$                 & \color{ForestGreen} \bf{58.7}  & \color{ForestGreen}  \bf{74.6} & \color{ForestGreen}  \bf{82.3}  \\
\hline
    S1                               & $95.6$                 & $90.8$                 & $27.3$                 & $66.1$                 & $76.9$                  \\
    S2                               & $96.0$                 & \color{ForestGreen} \bf{93.9}   & $42.3$        & $71.6$                 & $80.0$                  \\
    S3                               & $95.6$                 & $90.7$                 & $27.4$                 & $66.1$                 & $76.9$                  \\
    S4                               & \color{ForestGreen} \bf{96.2} & $92.6$          & $37.6$                 & $70.0$                 & $80.6$                  \\
%    \bottomrule
  \end{tabular}
\end{table}
\vspace{-0.3cm}
%\vspace{0.8cm}
\section{Conclusions}
De-identification of data used for ASR modeling poses a challenge for those tokens which become strongly suppressed in the training material, like names or dates. 
We propose various related strategies which are based on replacing PII tokens with surrogate tokens followed by generating corresponding audio via TTS or splicing and thereby 
enhancing the recognition performance of PII-related tokens compared to baseline methods. 
We evaluate the effectiveness of these strategies using a state-of-the-art conformer-transducer model architecture, 
comparing results to a best-case ``identified data'' baseline and models trained with de-identified data without any mitigation as worst-case baseline.

Results show that the best strategies can almost entirely recover the loss in general performance (WER) while maintaining a strong diarization performance (WDER). 
Most importantly, between $50\% - 90\%$ of the recall and precision drop in recognizing PII-related tokens can be recovered. While the splicing models perform better in 
terms of WER recovery, the TTS models show the highest F1 scores for PII-like tokens like names or other words strongly suppressed by de-identification. 

Future efforts will be based on trying to combine the splicing and TTS strategies, i.e., by using splicing if a matching word can be found in the corpus and otherwise 
falling back to TTS. We will also look for potential improvements by generating speech which more closely resembles the style surrounding the de-identified part of the data, 
both in terms of speaker characteristics (like accent, speaking rate, voice)  and environmental conditions (like background noise and reverberation).

\clearpage

\bibliographystyle{IEEEtran}

\bibliography{deid_paper}

\end{document}